# Double-layer Thin-film LiNbO$_3$ Longitudinally Excited Shear Wave Resonators with Ultra-large Electromechanical Coupling Coefficient and Spurious-Free Performance

Zhen-Hui Qin[1], Shu-Mao Wu[1], Chen-Bei Hao[1], Hua-Yang Chen[1], Sheng-Nan Liang[1], Si-Yuan Yu[1,2]* and Yan-Feng Chen[1,2]*

[1]National Laboratory of Solid-State Microstructures, Nanjing University, Nanjing 210093, China.
[2]College of Engineering and Applied Science, Nanjing University, Nanjing 210093, China.
corresponding authors*: Si-Yuan Yu (yusiyuan@nju.edu.cn), and Yan-Feng Chen (yfchen@nju.edu.cn)

**Abstract：** This work proposes a double-layer thin-film LiNbO$_3$ longitudinally excited shear wave resonator with a theoretical electromechanical coupling coefficient exceeding 60%, resonance-antiresonance (RaR) close to 28%, and no spurious modes. This ultra-large electromechanical coupling coefficient, which is close to the upper limit of LiNbO$_3$, is much larger than all microwave acoustic resonators reported so far. Based on X-cut thin-film LiNbO$_3$, when the film thickness is in the order of hundreds of nanometers, the frequency of the fundamental mode of the resonator can cover 1GHz to10GHz. The resonator design is convenient and flexible. The resonant frequency can be modulated monotonically by changing either the electrode or the thickness of the thin-film LiNbO$_3$ without introducing additional spurious modes. This ideal resonator architecture is also applicable to LiTaO$_3$. With the development of the new generation of mobile communications, this resonator is expected to become a key solution for future high-performance, ultra-wide-bandwidth acoustic filters.

**Introduction**

The development of 5G/6G mobile communications and sensing industries has put forward higher performance requirements for RF filters, including high frequency, large bandwidth, low insertion loss, high out-of-band suppression, high power tolerance, temperature stability, mechanical stability, and portability. Microwave acoustic devices are a type of device that convert microwave (electromagnetic) signals into acoustic signals through the piezoelectric effect, process the acoustic signals, and then convert them into microwave signals for output. Since the wavelength of acoustic waves is much shorter than electromagnetic waves, microwave acoustic devices are one of the important solutions to realize portable, high-performance RF filters. Currently, one of the crucial goals of microwave acoustic filters is to achieve large bandwidth under low insertion loss. The bandwidth requirements for RF filters in 4G communications are generally less than 10%, but the bandwidth of RF filters in future communications such as 6G may need to exceed 20%, which poses a great challenge to microwave acoustic filters.

Microwave acoustic filters can currently be divided into delay line and resonant filters. For a delay line filter, the bandwidth and insertion loss tend to change simultaneously [1,2]. For resonant filters, the bandwidth depends on the electromechanical coupling coefficients of resonators that compose the filter, while the insertion loss depends on the Q-factors of the resonators. Therefore, resonators are the key to microwave acoustic filters. The electromechanical coupling coefficient of an acoustic resonator mainly depends on the piezoelectric material used in the resonator, the acoustic mode, and the specific device structure. Currently, commonly used piezoelectric materials to construct microwave acoustic filters include $LiNbO_3$, $LiTaO_3$, AlN, AlScN, $SiO_2$, etc. Among them, $LiNbO_3$ has a relatively large electromechanical coupling coefficient and a relatively mature single-crystal preparation and processing method, which makes acoustic resonators based on $LiNbO_3$ an ideal component for realizing large-bandwidth filters. $LiNbO_3$ acoustic resonator first appeared in the late 1960s [3]. However, since these resonators were based on the Rayleigh mode (the electromechanical coupling coefficient was relatively small) at that time, people could not use them to make large-bandwidth microwave acoustic filters. In recent years, with the maturity of single-crystal thin-film $LiNbO_3$, a series of microwave acoustic resonators and filters based on thin-film $LiNbO_3$ have been proposed. They support more types of acoustic modes, such as A-modes [4-7] and S-modes [8-10] in Lamb waves, horizontal shear modes based on suspended films [11,12] and solid-mounted substrates

[13-15], etc. Many of these modes have electromechanical coupling coefficients much larger than the Rayleigh mode.

In 2019, Zhou et al. proposed a $SH_0$ mode resonator based on 30Y-X cut thin-film $LiNbO_3$ with an electromechanical coupling coefficient as high as 55% [16]. In 2020, Hsu et al. proposed a Love mode resonator based on X-cut thin-film $LiNbO_3$ with an electromechanical coupling coefficient reaching 44% [17]. In the same year, Lu et al. proposed an $A_1$ mode resonator based on 128Y-X cut thin-film $LiNbO_3$ with an electromechanical coupling coefficient exceeding 46% [18]. All these large electromechanical coupling coefficient resonators are based on transversely excited acoustic modes, so their electric field utilization is still relatively insufficient. To obtain higher electric field utilization, longitudinally excited resonators based on thin-film $LiNbO_3$ were subsequently proposed. In 2020, Plessky et al. proposed a longitudinally excited shear wave resonator in Y-cut thin-film $LiNbO_3$ (*i.e.*, YBAR), which is expected to achieve an electromechanical coupling coefficient of about 50% and a 22% resonance-antiresonance (RaR) [19]. In 2024, Hartmann et al. realized another longitudinally excited shear wave resonator with an electromechanical coupling coefficient exceeding 40% in an X-cut thin-film $LiNbO_3$ [20]. Although the above resonators have large electromechanical coupling coefficients, there are currently no solutions that can reach the theoretical upper limit of the $LiNbO_3$. In addition, spurious modes are prevalent in these large electromechanical coupling coefficient resonators [4,17,19,20,21], limiting their practical application in large-bandwidth RF filters.

In this paper, we propose a longitudinally excited shear bulk wave resonator based on a double-layer thin-film $LiNbO_3$. The structure of the resonator is shown in Figure 1, which consists of two bonded layers of thin-film $LiNbO_3$, as well as a top and a bottom electrode above and below them, respectively. Under this design, the electromechanical coupling coefficient of the resonator can reach 60.5%, and the RaR reaches 27.9%, much higher than all microwave acoustic resonators reported so far. The longitudinal excitation of the acoustic mode in the thin-film $LiNbO_3$ brings the electric field utilization close to 100%, ensuring that the resonator's electromechanical coupling coefficient is close to the theoretical maximum of the $LiNbO_3$. The significance of the double-layer structure is that it achieves perfect suppression of spurious modes in the primary mode's band, without affecting the electromechanical coupling coefficient of the resonator. Also, in this resonator, the resonant frequency can be easily modulated by the thickness of any single-layer electrode or single-layer thin-film $LiNbO_3$, without introducing any additional spurious modes. This highly ideal acoustic resonator may become a key solution in future ultra-wide-bandwidth mobile communications.

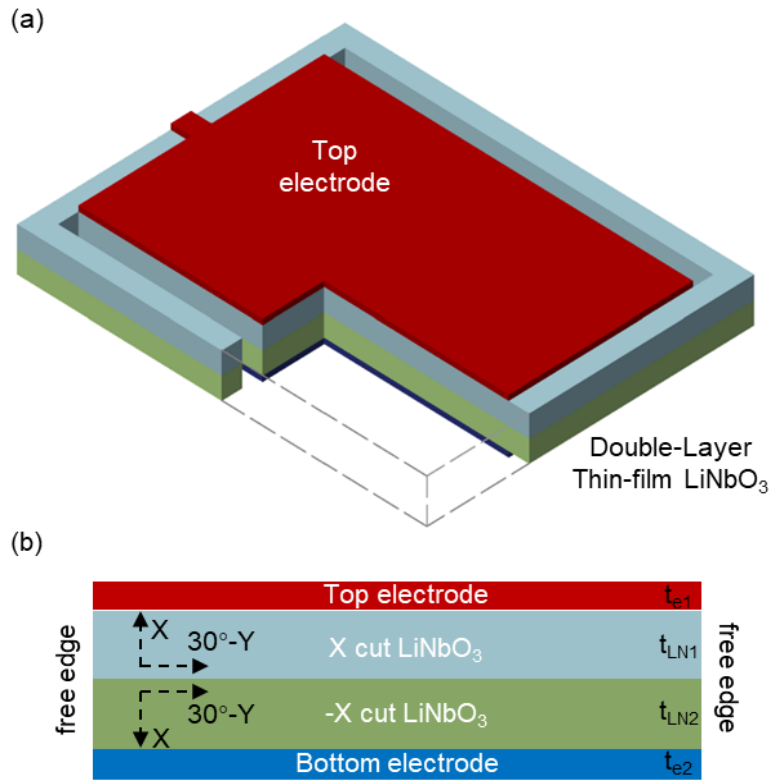

**Figure 1**. Structural diagram of the double-layer thin-film LiNbO$_3$ shear wave resonator

(a) Three-dimensional diagram (b) Cross-sectional view

**Single-layer LiNbO$_3$ thin-film resonator**

The electromechanical coupling coefficient of a piezoelectric material mainly depends on its piezoelectric coefficient. Figure 2 shows the relationship between the piezoelectric coefficient $e_{34}$ of the longitudinally excited shear wave and the horizontal rotation angle of the crystal (perpendicular to the amplitude direction of the shear wave) in the common cuts of LiNbO$_3$. For $e_{34}$, X-cut LiNbO$_3$ has significant advantages over other cuts. Therefore, we choose X-cut thin-film LiNbO$_3$ to construct longitudinally excited shear wave resonators.

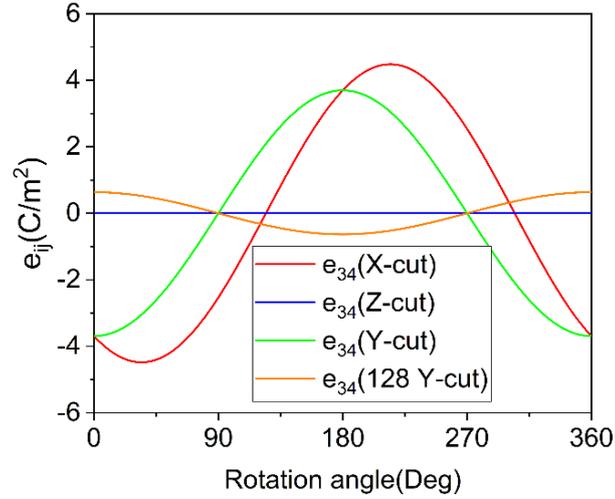

**Figure 2.** Among the four common thin-films LiNbO$_3$ (X-cut, Z-cut, Y-cut, and 128-Y-cut), the piezoelectric coefficient corresponding to the longitudinally excited shear mode, *i.e.*, e$_{34}$, under different rotation angle of the LiNbO$_3$ crystal.

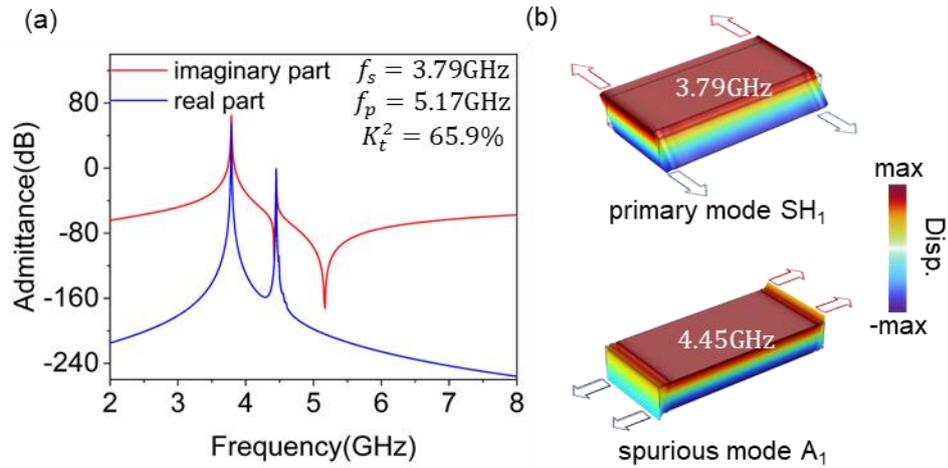

**Figure 3.** (a) Admittance spectra of single-layer thin-film LiNbO$_3$ longitudinally excited shear wave resonator (b) Displacement field distributions of its primary and spurious modes, belonging to SH$_1$ and A$_1$ modes, respectively.

We take 300nm X-cut thin-film LiNbO$_3$ as an example to build a resonator model. Gold (Au) electrodes with a thickness of 20 nm are provided on the upper and lower sides of the thin-film LiNbO$_3$. The three-dimensional (3D) simulation results of the resonator are shown in Figure 3. Figure 3 (a) is the admittance spectra of the resonator (real and imaginary parts). Figure 3 (b) shows the displacement fields of the primary and spurious modes. The results show that the primary mode of the resonator is SH$_1$, located at 3.79GHz, and has an electromechanical coupling coefficient as high as 65.9%. However, there is a significant spurious mode within the passband of this primary mode, located at 4.45GHz. The spurious mode belongs to the A$_1$ mode, the corresponding piezoelectric coefficient is e$_{35}$, and has the same electrical excitation direction and mode order as the primary mode. This means that it is difficult to suppress this spurious mode using traditional solutions, while the primary mode maintains the existing electromechanical coupling coefficient.

**Double-layer LiNbO$_3$ thin-film resonator**

The spurious mode described above makes the thin-film LiNbO$_3$ resonator unusable for building filters. To suppress such spurious mode, we innovatively use a double-layer thin-film LiNbO$_3$ instead of the single-layer thin-film. This solution completely eliminates unwanted spurious mode while maintaining the resonator's ultra-large electromechanical coupling coefficient.

Theoretically, the piezoelectric coupling equation in piezoelectric crystals can be written as.

$$\begin{cases} T = c\dfrac{\partial u}{\partial x} - e\dfrac{\partial \varphi}{\partial x} \\ D = e\dfrac{\partial u}{\partial x} + \varepsilon\dfrac{\partial \varphi}{\partial x} \end{cases} \quad (1)$$

The bulk acoustic wave resonator based on a single-layer thin-film LiNbO$_3$ can be treated as a one-dimensional (1D) model. Ignoring the influence of the electrode, the admittance spectrum of the resonator can be approximated, as:

$$Y \propto 1 / \left(1 - \dfrac{k_t^2 \tan\left(\dfrac{\pi f d}{v}\right)}{\dfrac{\pi f d}{v}}\right) \quad (2)$$

Among them, $k_t^2 = \dfrac{e_{43}^2}{\varepsilon_{33} c_{44} + e_{43}^2}$ is the electromechanical coupling coefficient, $f$ is the frequency, $d$ is the thickness of the thin-film, and $v$ is the acoustic velocity of the specific mode. When the upper and lower boundaries of the thin-film are free, the high-order modes usually share the same symmetry (high-order modes all belongs to either S mode or A mode) [22]. The resonance condition of the modes is

$$\dfrac{\pi f d}{v} = n\pi + \dfrac{\pi}{2} \quad (3)$$

For a bulk acoustic wave resonator based on a double-layer, equal-thickness ($d$) thin-film LiNbO$_3$, two situations can be simply considered. The first is that the piezoelectric coefficients of the upper and lower LiNbO$_3$ films are the same. The second is that the piezoelectric coefficients of the upper and lower LiNbO$_3$ films have the same magnitude but in opposite directions.

For the first situation, the admittance spectrum of the resonator can be directly described by Formula (2). At this time, the resonance conditions of the resonator are the same as a single-layer thin-film with a thickness of $2d$. The acoustic fields in the upper and lower thin-films are antisymmetric along the film thickness direction.

$$\frac{2\pi f d}{v} = n\pi + \frac{\pi}{2} \qquad (4)$$

For the second situation, the piezoelectric coefficient is reversed but the electric field direction remains unchanged, which can be equivalent to the piezoelectric coefficient remaining unchanged but the electric field direction is reversed. That is to say (-e)·E=(-E)·e. Therefore, the acoustic fields in the upper and lower thin-films are symmetric along the film thickness direction. At this time, the resonance conditions of the double-layer resonator are the same as those of a single-layer resonator with thickness $d$.

$$\frac{\pi f d}{v} = n\pi + \frac{\pi}{2} \qquad (5)$$

As mentioned above, the piezoelectric coefficient e$_{34}$ corresponding to the longitudinally excited shear wave resonator has the maximum value in the X-cut LiNbO$_3$. Therefore, we also choose X-cut thin-film LiNbO$_3$ film to construct double-layer longitudinally excited shear wave resonators. Figure 4 shows the piezoelectric coefficients of the primary and spurious modes in X-cut LiNbO$_3$ and -X-cut LiNbO$_3$. The primary and spurious modes are SH$_2$ and A$_1$, corresponding to piezoelectric coefficients e$_{34}$ and e$_{35}$, respectively. e$_{34}$ has the same value but opposite sign in X-cut LiNbO$_3$ and -X-cut LiNbO$_3$; e$_{35}$ has the same value and sign in X-cut LiNbO$_3$ and -X-cut LiNbO$_3$. Therefore, the SH$_2$ mode belongs to the second resonance condition mentioned above, while the spurious A$_1$ mode belongs to the first resonance condition. Figure 5 shows the vibration forms of these two modes in single-layer and double-layer thin-film LiNbO$_3$ resonators. According to formulas (4) and (5), in the double-layer thin-film LiNbO$_3$ resonator, the frequency of the spurious A$_1$ mode will be far away from the passband of the primary SH$_2$ mode, thereby eliminating the disadvantages caused by the spurious mode.

Note that to ensure that the double-layer thin-film LiNbO$_3$ resonator has the maximum electromechanical coupling coefficient, we make the 30-Y direction of the upper and lower thin-film LiNbO$_3$ consistent, so that the piezoelectric coefficient e$_{34}$ of their primary modes reaches the maximum value.

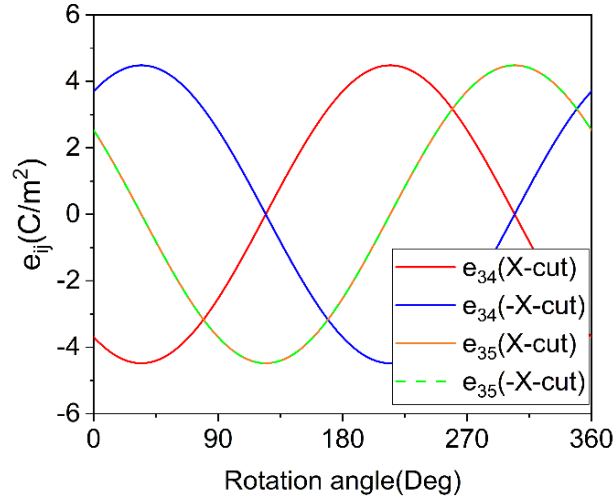

**Figure 4.** In X-cut and -X-cut thin-film LiNbO$_3$, the piezoelectric coefficients corresponding to the primary SH$_2$ and spurious A$_1$ modes of the resonator, *i.e.*, e$_{34}$ and e$_{35}$, under different rotation angle of the LiNbO$_3$ crystal.

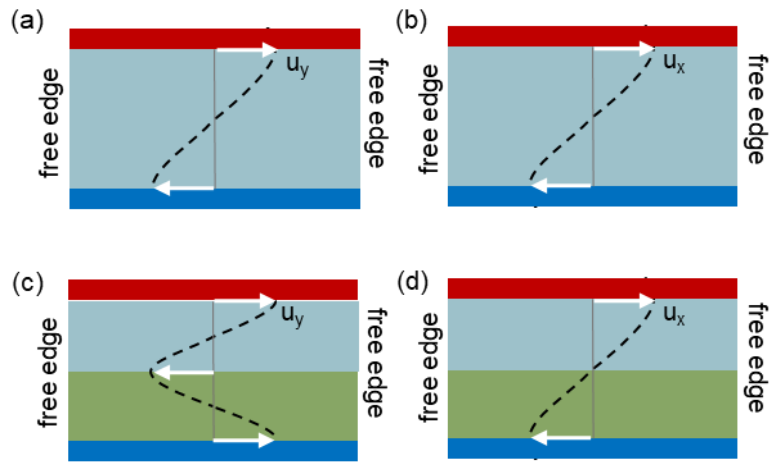

**Figure 5.** Resonance modes in single-layer and double-layer thin-film LiNbO$_3$ resonators (a) Single-layer thin-film LiNbO$_3$, primary SH$_1$ mode (b) Single-layer thin-film LiNbO$_3$, spurious A$_1$ mode (c) Double-layer thin-film LiNbO$_3$, primary SH$_2$ mode (d) Double-layer thin-film LiNbO$_3$, spurious A$_1$ mode.

We take a double-layer thin-film composed of 300nm X-cut LiNbO$_3$ bonded with 300nm -X-cut LiNbO$_3$ as an example to build a resonator model. Consistently, Au electrodes with a thickness of 20 nm are provided on the upper and lower sides of the double-layer thin-film LiNbO$_3$. The 3D simulation results of the resonator are shown in Figure 6. Figure 6 (a) is the admittance spectra of the resonator (real and imaginary parts). Figure 6 (b) shows the displacement fields of the primary and spurious modes. The results show that the primary mode of the resonator is SH$_2$, located at 4.71GHz, and has an electromechanical coupling coefficient of 60.5%. The A$_1$ spurious mode is at 2.68GHz, much lower than the primary mode, and its intensity is also faint. Clearly, in the double-layer thin-film LiNbO$_3$ resonator, we successfully eliminated the adverse effects of the spurious mode while ensuring that the primary mode still has a considerable electromechanical coupling coefficient. The RaR of the primary mode of the resonator reaches an astonishing 27.9%, significantly higher than all reported microwave acoustic resonators.

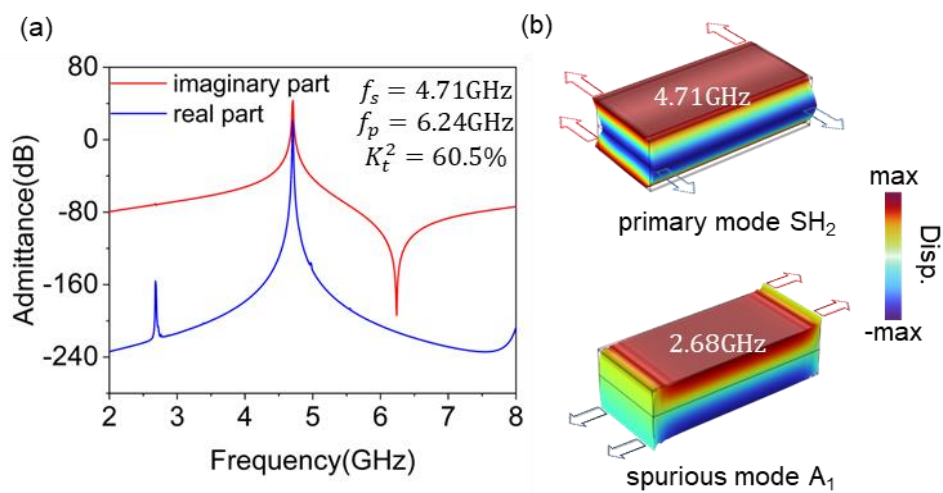

**Figure 6.** (a) Admittance spectra of double-layer thin-film LiNbO$_3$ longitudinally excited shear wave resonator (b) Displacement field distributions of its primary and spurious modes, belonging to SH$_2$ and A$_1$ modes, respectively.

**Resonator design: applicable and convenient**

Our proposed double-layer thin-film LiNbO$_3$ longitudinally excited shear wave resonator also has advantages in terms of design applicability and convenience. The operating frequency of the resonator changes monotonically with the thickness of the thin-film LiNbO$_3$ and electrodes. When the main design parameters of the resonator (*e.g.*, thickness of the thin-film LiNbO$_3$, thickness of the top/bottom electrodes, orientation of the X-cut LiNbO$_3$) are changed, its electromechanical coupling coefficient can basically be maintained at a considerable value, and almost no spurious modes will be introduced.

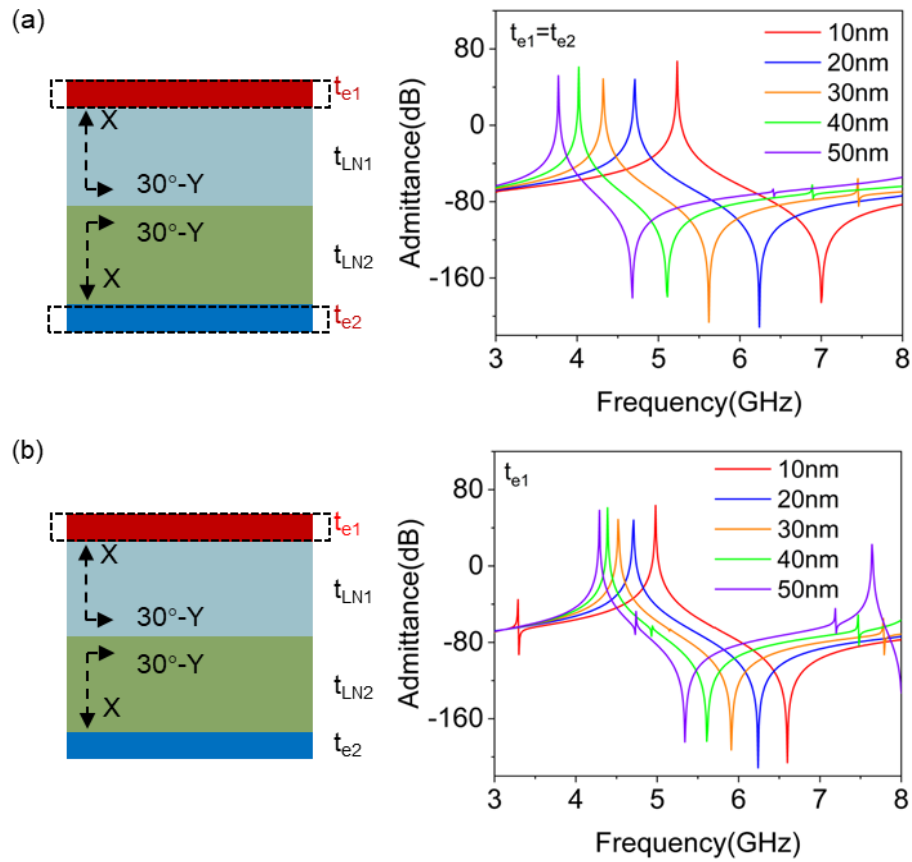

**Figure 7.** Admittance spectra of the resonators with different top and bottom electrode thicknesses (a) The thickness of the top and bottom electrodes changes simultaneously (b) The thickness of the bottom electrode remains unchanged at 20nm, while the thickness of the top electrode changes

Figure 7 shows the admittance spectra of the resonators with different electrode thicknesses. Figure 7(a) corresponds to the case of changing the thickness of the top and bottom electrodes simultaneously. The thinner the electrode, the higher the resonant frequency and the greater the electromechanical coupling coefficient. However, when the electrode thickness is as high as 50nm, the resonator still has an electromechanical coupling coefficient greater than 50%. In all cases, there is a complete absence of spurious modes within the passband range of the resonator's primary mode. Figure 7(b) shows the admittance spectra under different top electrode thicknesses when the bottom electrode thickness is constant at 20nm. The resonant frequency and electromechanical coupling coefficient decrease as the thickness of the top electrode increases. Only when the thickness difference between the top and bottom electrodes is quite large, a weak spurious mode will appear in the passband of the primary mode, which is not as ideal as the case in Figure 7(a).

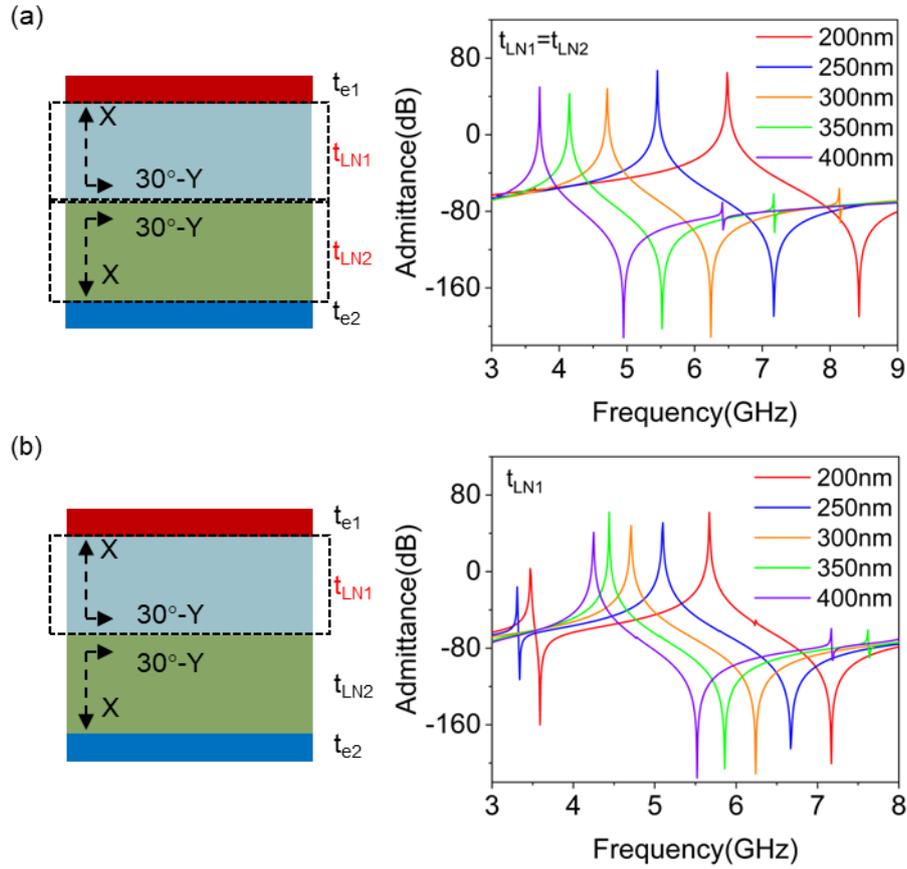

**Figure 8.** Admittance spectra of the resonators under different thin-film LiNbO$_3$ thicknesses (a) The thickness of the upper and lower thin-film LiNbO$_3$ changes simultaneously (b) The thickness of the lower thin-film LiNbO$_3$ is fixed at 300nm, while the thickness of the upper thin-film LiNbO$_3$ changes.

Figure 8 shows the admittance spectra of the resonators under different thin-film LiNbO$_3$ thicknesses. Figure 8(a) corresponds to the case of changing the thickness of the upper and lower thin-film LiNbO$_3$ simultaneously. The thinner the thin-film LiNbO$_3$, the higher the resonant frequency, while the electromechanical coupling coefficient almost unchanged. When the thickness of the thin-film LiNbO$_3$ changes from 200nm to 400nm, the resonant frequency changes quasi-linearly from 3.7GHz to 6.8GHz. In all cases, there is a complete absence of spurious modes within the passband range of the resonator's primary mode.

It needs to be taken into account that the filter needs to contain two sets of resonators with different frequencies. However, during actual device processing, it is extremely difficult to change the thickness of the upper and lower thin-film LiNbO3 simultaneously in a single wafer. If the frequency of the resonator can be changed only by changing the thickness of the upper thin-film LiNbO3 (for example, by regional thinning method [23]), it will be highly beneficial to the implementation of the filter. Figure 8(b) shows the resonator admittance spectra under different upper thin-film LiNbO3 thicknesses when the thickness of the lower thin-film LiNbO3 is fixed at 300nm. At this time, the resonant frequency of the resonator also increases quasi-linearly as the upper thin-film LiNbO3 becomes thinner, with the electromechanical coupling coefficient hardly changing (always greater than 56%). In all cases, no significant spurious modes appear in the passband of the resonator's primary mode. This means the double-layer LiNbO3 film resonator is practical and feasible in composing ultra-large bandwidth filters.

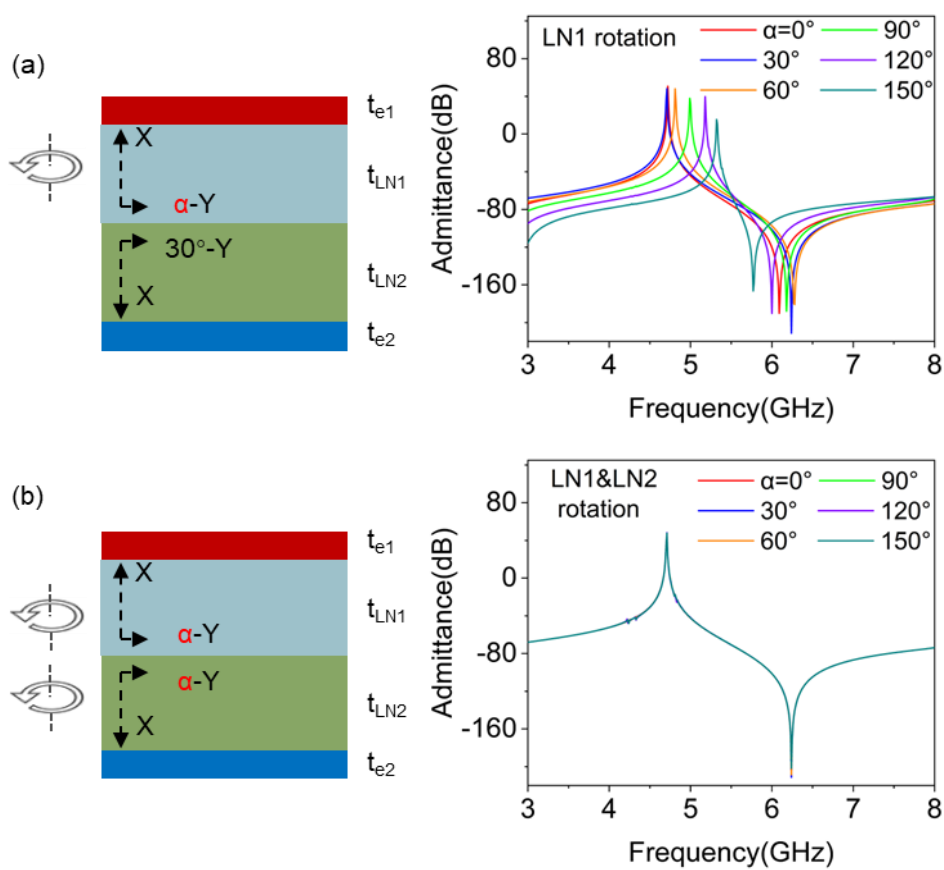

**Figure 9.** Admittance spectra of the resonators under different orientation of the X-cut thin-film LiNbO3 (a) The rotation angle of the lower X-cut thin-film LiNbO3 is fixed at 30°-Y, while the rotation angle of the upper X-cut thin-film LiNbO3 changes. (b) The rotation angle of the upper and lower X-cut thin-film LiNbO3 changes simultaneously.

Figure 9 shows the admittance spectra of the resonators under different rotation angle of the X-cut thin-film LiNbO$_3$. Figure 9 (a) shows the case only the rotation angle of the upper thin-film LiNbO$_3$ is changed. The electromechanical coupling coefficient decreases monotonically as the upper X-cut thin-film LiNbO$_3$ deviates from its original rotation angle (*i.e.*, 30°-Y). What is noteworthy and advantageous is the simultaneous change of the rotation angle of the upper and lower thin-film LiNbO$_3$, as shown in Figure 9(b). At this time, no matter how much the two X-cut thin-film LiNbO$_3$ deviate from their original rotation angle, as long as the two thin-films are synchronized (*i.e.*, their 30°-Y directions coincide), the admittance spectra of the resonators will be completely consistent, all providing ultra-large electromechanical coupling coefficient and near-perfect spurious mode suppression. It can be seen that the performance of this type of resonator mainly depends on the quality of the double-layer thin-film LiNbO$_3$, but it has a high tolerance for electrode design and device processing accuracy.

**Conclusion**

This paper proposed a longitudinally excited shear wave resonator based on a double-layer X-cut thin-film LiNbO$_3$, which can provide an electromechanical coupling coefficient of up to 60.5%, an RaR of up to 27.9%, and near perfect spurious-free performance. The resonant frequency of this resonator mainly depends on the thickness of the thin-film LiNbO$_3$ rather than the electrode line width (*i.e.*, photolithography capability). With the thin-film LiNbO$_3$ thickness on the order of hundreds of nanometers, the operating frequency of the resonator can cover 1GHz to 10GHz. The resonator's operating frequency and electromechanical coupling coefficient can be flexibly designed by changing the thickness/material of the top electrode, the thickness of the upper thin-film LiNbO$_3$ or rotating the upper thin-film LiNbO$_3$, which is of great significance in realizing ultra-large bandwidth RF filters. At the same time, this suspended thin-film LiNbO$_3$ resonator does not require any anchor structures, it therefore will have higher mechanical and temperature stability than other suspended thin-film LiNbO$_3$ resonators. Although LiNbO$_3$ is used as a demonstration in this paper, the scheme is also applicable to LiTaO$_3$ with the same symmetry but higher temperature stability [24]. This double-layer thin-film bulk acoustic resonator, which brings the electromechanical coupling coefficient close to the theoretical upper limit of the material, is expected to play a key role in future mobile communications with ultra-large bandwidth and ultra-high-performance.